\documentclass{aastex63}

\received{\today}
\revised{\today}
\accepted{\today}
\submitjournal{ApJ}

\shorttitle{Detailed Study of the Sunspot Group's Decay }
\shortauthors{J. Murak\"ozy}

\begin{document}

\title{On the Decay of Sunspot Groups and Their Internal Parts in Detail}

\correspondingauthor{Judit Murak\"ozy}
\email{murakozy.judit@csfk.mta.hu}

\author[0000-0001-6920-259X]{Judit Murak\"ozy}
\affiliation{Konkoly Thege Mikl\'os  Astronomical Institute\\ Research Centre for Astronomy and Earth Sciences\\
Konkoly-Thege Mikl\'os \'ut 15-17.\\
Budapest, 1121, Hungary}


\begin{abstract}
The decay of sunspot groups is a relatively unknown field since most studies have focused mainly on the decay of sunspots or sunspot groups, but only on small samples. As an extension of the recent work of \citet{2020ApJ...892..107M} which is based on a large verified sample, this study investigates not only the long-term behavior of the decay of sunspot groups but also the dynamics of their parts. The aim of the present work is to search for dependencies of the decay process in order to find physical conditions that modify or contribute to the decay. The investigations are based on the catalog of SoHO Debrecen Sunspot Database (SDD ) and the Greenwich Photoheliographic Results (GPR) as well as the Debrecen Photoheliographic Data (DPD). Altogether more than 750 sunspot groups were considered. The decay rates have been calculated for the total, umbral and penumbral area of the groups and in the case of the SDD's groups they have been calculated for both the leading and the following parts. The decay rates depend linearly on the maximum areas and ranged from 30-50 MSH/day for the sunspot groups and penumbrae and 5-10 MSH/day for the umbrae throughout the cycle. The decay rates fall significantly during the Gnevyshev gap and show 4+4 Schwabe cyclical variations in the ascending/descending phases, but it is always higher in the northern hemisphere. There is a slight decrease of the decay rates in the activity range toward higher latitudes.
\end{abstract}

\section{Introduction}
           \label{S-Introduction}

The decay of the solar active regions and their morphological parts is poorly investigated phenomenon. The published investigations are mainly based on small samples, the long term studies focus only on the decay of total sunspot groups, their umbra/penumbra ratio or sunspots.
\cite{2018ApJ...865...88C} investigated 48 different sunspots during the Maunder Minimum and pointed out a relationship between the umbra/penumbra (U/P) ratio and the sunspot decay. They found that higher U/P values mean faster sunspot decay. \citet{1993A&A...274..521M} and \citet{1993ASPC...46...67M} showed that the U/P ratio is constant during the decay and the decay rate varies by four solar cycles. \citet{1982SoPh...81...25R} studied 31 different U and P areas and obtained smaller decay rates for greater areas and pointed out that larger spots are short-lived. \cite{2008SoPh..250..269H} studied active regions from 130 years observations and found that high latitude spots decay more rapidly than the low latitude ones and the decay rate is higher at the maximum and on the raising phase of the cycle while it is lower on the declining phase and at the cycle minimum. They also showed that the decay speeds of sunspot groups depend on their areas; the higher the area or percentage area the higher the decay rate. \cite{1963BAICz..14...91B} showed that the decreasing phase of sunspot groups area begins with a faster decay followed by a slower regression, furthermore, long-lived sunspot groups decay more slowly.
\citet{1995SoPh..157..389L} investigated the growth and decay of sunspot groups and pointed out that the daily decrease of the latitudes depends on the latitude of sunspot groups.
\cite{2011SoPh..270..463J} studied the growth and decay rates of sunspots on a long time scale and reported that these rates are symmetric to the equator and the decay rates vary by 13\% on a 60-year time scale. He pointed out that the decay rates were the lowest of the past 100 years near the end of Cycle 23, but at the beginning of Cycle 24 they were the largest. He also found that the north-south asymmetry in the growth rates of sunspot groups has a modulation of 33-44 years as well as the area decrease is higher near the minimum of the solar cycles and lower around the cycle maximum.\\
Some studies of decays that investigated the leading and following parts separately can also be found in the literature. \citet{1992SoPh..142...47H} investigated the motions of plages and sunspot groups at their development and decaying phases by using Mt. Wilson data. He found that during the decay the leading part rotates slower by about 3\% than the following part of the sunspot group furthermore the decaying group rotates significantly faster than the growing group. 
A recent study \citep{2020ApJ...892..107M} investigated the decay of only the umbral area of sunspot groups using SoHO/MDI data which make possible to treat the leading and following parts of active regions separately. As a result of the study of 206 different active regions one can see that the decay rate of following parts is higher than that of the leading parts. This study did not take into account any long term effect or hemispheric differences. As \citet{2012MNRAS.419.3624M} and \citet{2016ApJ...826..145M} reported a characteristic 4+4 pattern of hemispheric cycles in which the cycle of Northern hemisphere leads in four Schwabe cycles and the Southern hemispheric cycle leads in the next four cycles. The cause of this pattern is yet unclear. The present paper also searches any sign of the mentioned 4+4 alternation in the decay rates.

To interpret the decay process \citet{2013MmSAI..84..428G} used a numerical model to investigate the sunspot decay and found that the decay could be explained by advection and stretching of the magnetic field caused by a negative radial velocity field, however their sample contains only four active regions.\\
\citet{2014ApJ...785...90R} investigates the process of sunspot formation and decay by using numerical simulation. They obtained that the decay of sunspots follows the formation immediately in the trailing part but there is a short stationary state between the two phases in the leading part. During the emergence of the magnetic flux into the photosphere the torus-aligned flow in the convection zone forms the flux ropes and makes more axisymmetric and coherent leading spots and more fragmented following spots. In their simulations the decay is driven by the largest scale convective motions but the obtained timescale is as short as a few days. \citet{1974MNRAS.169...35M} suggested that the decay of sunspots is related to the supergranular convection by using model study.

The decaying sunspot is considered as a decreasing area after the time of maximum area and this decrease may governed by several mechanisms. One of them is the turbulent erosion which carried out the magnetic flux from the sunspot perimeter. 
\citet{1997SoPh..176..249P} found that the decay rates are proportional to the square root of the sunspot area and they explained this with the impact of the so-called turbulent erosion on the perimeter of the sunspot. 
Observational report of \citet{2003GeoRL..30.1178C} have analyzed 32 sunspots and they have not found significant correlation between the decay rate and the square root of the sunspot area. They concluded that the decay occurs throughout the whole sunspot and not just at its edge. In addition they observed that the larger the U/UP area, the smaller the decay rate because the larger umbra somehow stabilizes the sunspot against the decay.
The other possible procedure governed by the moving magnetic features (MMFs) which are small knots around the sunspots with roughly 10$^3$ km diameter. These magnetic fluxes move outward from the spots with a speed of about 1 km/s.
\citet{2005ApJ...635..659H} studied these small-scale MMFs around eight sunspots and found more MMFs around larger spots and the rate of the net flux carried away by them from a spot is (0.4-6.2)x10$^{19}$ Mx hr$^{-1}$ if the sunspot decay is governed only by the outflowing MMFs.
Recently \citet{2017ApJ...842....3N} studied 10 sunspot groups and followed their leading and following parts separately. They found that the following polarity  decays more rapidly and the decay rate of the group is in correlation with the rate of MMFs. The flux decay rate is proportional to 0.57th power of peak flux and its average 3.2x10$^{19}$ Mx/hr value is agreement with the mentioned rate of \citet{2005ApJ...635..659H}.
The relationship between the growth and decay and the convection upward and downward flow is also investigated on a sample of 788 ARs and found that during the decay the strong upflows changed to downflows at depths greater than 10 Mm, but there is no change at other depths \citep{2009SoPh..258...13K}.

The present investigation focuses on the details of the decaying phase of sunspot groups as well as their internal dynamics and their possible long term and hemispheric variations.

\section{Data}
	\label{data}
The input data are the sunspot catalogs made in the Debrecen Heliophysical Observatory. Two of them are based on ground-based observations (revised version of the Greenwich Photoheliographic Results (hereafter GPR) and Debrecen Photoheliographic Data (hereafter DPD) \cite{2016SoPh..291.3081B}), while the third data set, namely SoHO/MDI -- Debrecen Sun-spot Data (hereafter SDD) is made by using space-borne data observed by Solar and Heliospheric Observatory (SoHO) \cite{2015MNRAS.447.1857B}. The first two databases do not contain magnetic data, the opposite polarity parts of sunspot groups cannot be distinguished, but as long-term data sets they allow to study the possible cyclic dependence of any features. The SDD contains magnetic data for each sunspot thus it makes possible to distinguish between the leading and following parts.\\
The time resolution of the ground-based data sets is one day and that of the SDD is 1.5 hours. All databases contain the data of total as well as the umbral and penumbral areas of sunspot groups. The areas are measured in millionths of the solar hemisphere (MSH).

\section{Methods and samples} 
\begin{figure}[h!]
\center
\includegraphics[angle=0, width=14cm]{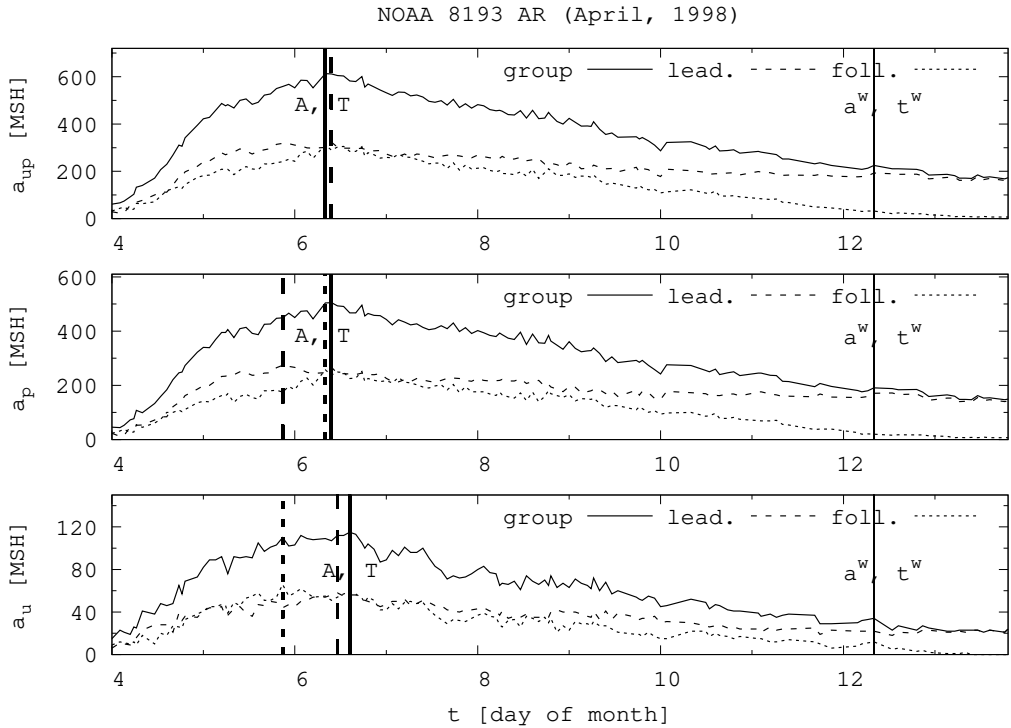}
\caption{Evolution of active region NOAA 8193 (April, 1998). The total, penumbral and umbral area data of the sunspot group are taken from the SDD. The thick solid/dashed/dotted vertical lines mark the times of maximum total/leading/following areas (T, A), while the solid thin vertical lines represent the time of the last observation before the central meridian distance of 65$^{\circ}$ (a$^w$, t$^w$).  In the case of a$_{up}$ the time of maximum area of the following part coincides with that of the total area maximum. By using Equation~\ref{eq:decay} the decay rates of this sunspot group can be found in the text.}
\label{fig:method}
\end{figure}

The sunspot decay has been investigated by using a relatively simple method. The descending phase of the sunspot group area is considered to have a linear variation between the maximum and final observed areas. Final area is measured at the last time when the sunspot group can be observed within 65 degrees measured from the central meridian (Fig.~\ref{fig:method}). The decay rate of the group's area is

\begin{equation}
      d = \frac{a^{w}-A}{t^{w}-T}
\label{eq:decay}
\end{equation}

where A is the maximum area measured at T, while $a^w$ is the area measured at the $t^w$ time which is the last observed time within the 65 degrees western central meridian distance. The results are expressed in MSH/day.
The percental decrease of the sunspot group's area, i.e., the percentage decay rate can be defined as
\begin{equation}
      d = \frac{(A-a^{w})/A}{t^{w}-T}
\label{eq:percdecay}
\end{equation}
in order to eliminate the area (the results will be obtained in \%/day).

Sunspots of different sizes have to be treated uniformly when determining the decay rate therefore it is calculated as a variation of the area normalized to the maximum area expressed in percentages.
This method is appropriate during only the first phase of the decay (as Fig.~\ref{fig:method} illustrates) where the decay speed is the highest. In the cases of recurrent groups this method does not give correct results because of the different decay speeds. Therefore the long-lived returning sunspot groups, e.g., AR NOAA 7031 are omitted from this study.

The selected sunspot groups vary in the following way. Very strict criteria are defined to identify the real decaying groups. The chosen sunspot groups have to be observable at least for eight days but their areas have to decrease during at least four days and have to increase during at least two consecutive days. This decrease has to be at least 60\% of the maximum area. These criteria ensure the identification of the real maximum area of sunspot groups.
It is also required that not only the total, i.e., umbra+penumbra area of sunspot groups has to decrease but also the umbral area. The penumbral area is a$_p$ = a$_{up}$-a$_u$ where a$_{up}$ means the total, i.e., umbra+penumbra area of sunspot groups corrected for foreshortening, and a$_u$ is the corrected umbral area. The sunspot groups have to be bipolar, i.e., they consist of leading and following portions at the maximum area state of sunspot groups. 

The studies based on the SDD can take a difference between the leading and following parts of sunspot groups and treat them as three entities. Therefore, in the case of one sunspot group we have nine areas i.e. its total area and its areas of the leading and following parts that can be obtained for the umbra+penumbra, penumbral and umbral areas. Thus, nine different maximum areas and times as well as data for the last state result nine different decay rates. In that case if the percentage area regression is taken into account as well, we have 18 decay rates for only one sunspot group. The time profiles of AR NOAA 8193 are plotted in Fig.~\ref{fig:method} as an example from the sample taken from the SDD. The calculated decay rates are as follows:\\
d$_{up}=(225-613)/6=-64.67$ MSH/day,\\
d$_{up}^l=(193-323)/5.93=-21.92 $ MSH/day, \hspace{3mm}
d$_{up}^f=(32-312)/6=-46.67 $ MSH/day \vspace{5mm} \\
d$_p=(191-505)/5.93=-52.95$ MSH/day, \\
d$_p^l=(171-274)/6.47=-15.92$ MSH/day, \hspace{3mm}
d$_p^f=(20-257)/6=-39.50$ MSH/day\vspace{5mm} \\ 
d$_u=(34-115)/5.73=-14.14$ MSH/day, \\
d$_u^l=(22-58)/5.87=-6.13$ MSH/day, \hspace{3mm}
d$_u^f=(12-66)/6.47=-8.35$ MSH/day.

\section{Results and Discussion}
\subsection{Decay Rates}
\begin{figure}[h!]
\center
\includegraphics[angle=0, width=0.85\linewidth]{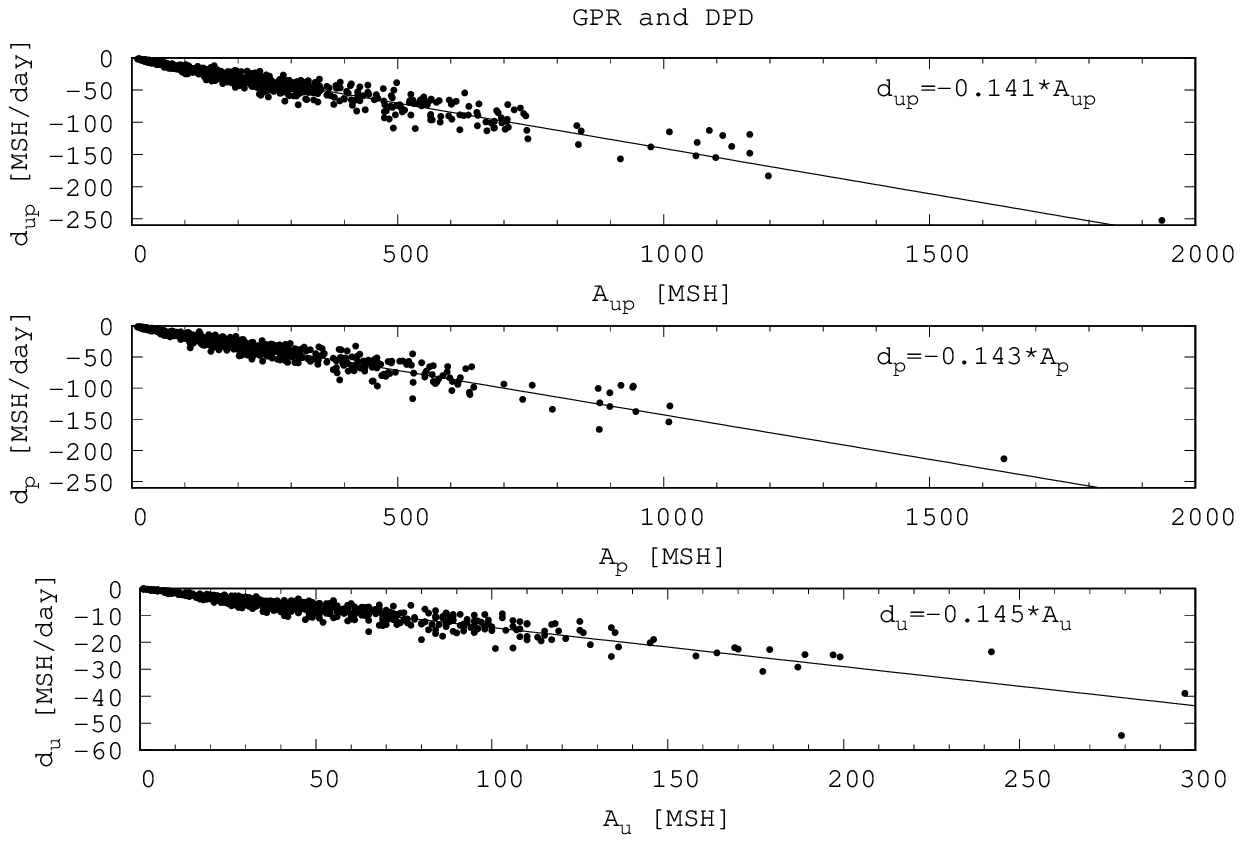}\\
\includegraphics[angle=0, width=0.85\linewidth]{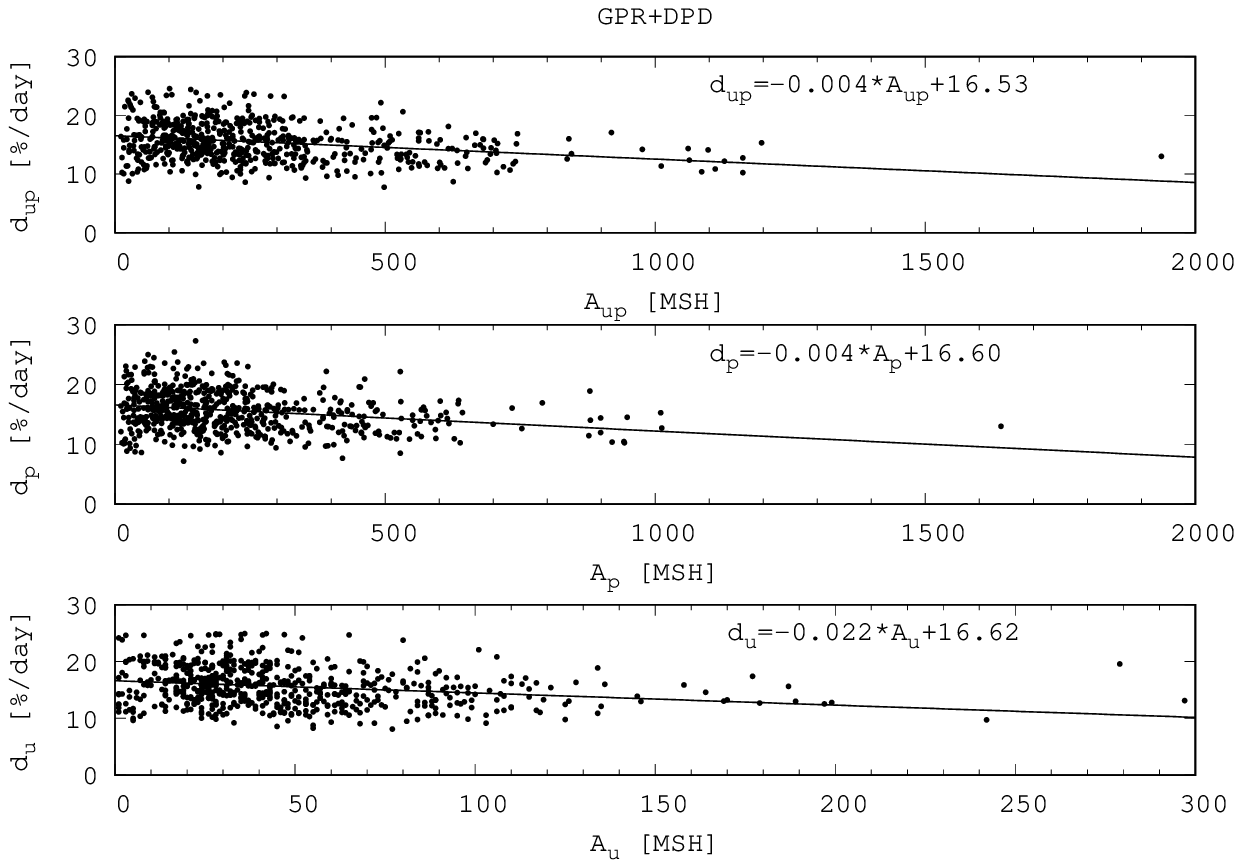}
\caption{Decay rates expressed in MSH/day (upper three panels, MSH means Millionths of Solar Hemisphere) and percentage decay rates expressed in \%/day (lower three panels) as a function of the proper maximum areas. Upper, middle and lower panels of the triplets refer to the whole area, penumbral area and umbral area respectively. The figures contain the data of all investigated sunspot groups from GPR and DPD. Each point marks a sunspot group.}
\label{fig:decrates}
\end{figure}

The decay rate of sunspot groups was calculated for the three areas, i.e., the total area, the penumbral area, and the umbral area by using the GPR (1874--1976) and DPD (1977--2018) catalogs as well as the SDD (1996--2010). The number of the investigated sunspot groups are 305 and 321 from the GPR and DPD, respectively. The investigated number of sunspot groups collected from SDD is 142. All of the decay rates of sunspot groups selected from GPR and DPD depend on the proper maximum area, the higher the areas the higher the decay rates (see the top panels of Fig.~\ref{fig:decrates}). \citet{2008SoPh..250..269H} found the same result. By using the GPR and DPD data sets the decay rates of the total and penumbral areas, as well as the umbral areas are almost equal.

The three bottom panels of Fig.~\ref{fig:decrates} also contain all the data collected from GPR and DPD and show a different view of the process. The decay rates are expressed in percentages of the maximum area and in contrast to the decay diagrams expressed in MSH (higher decay rates in larger areas) the percental decay rate of larger sunspot groups is somewhat lower than that of the small groups. Apparently, this may be responsible for the longer life of larger groups. This result is in contrast with that of \citet{2008SoPh..250..269H}, but \citet{1992SoPh..137...51H} has found a similar relationship.

The diagrams of Fig.~\ref{fig:decrates} may imply that the decay processes of umbrae and penumbrae are controlled by different mechanisms. The data do not belong to an individual spot but they are summarized for all umbrae and penumbrae in the active region.

\begin{figure}[h!]
\center
\includegraphics[angle=0, width=0.9\linewidth]{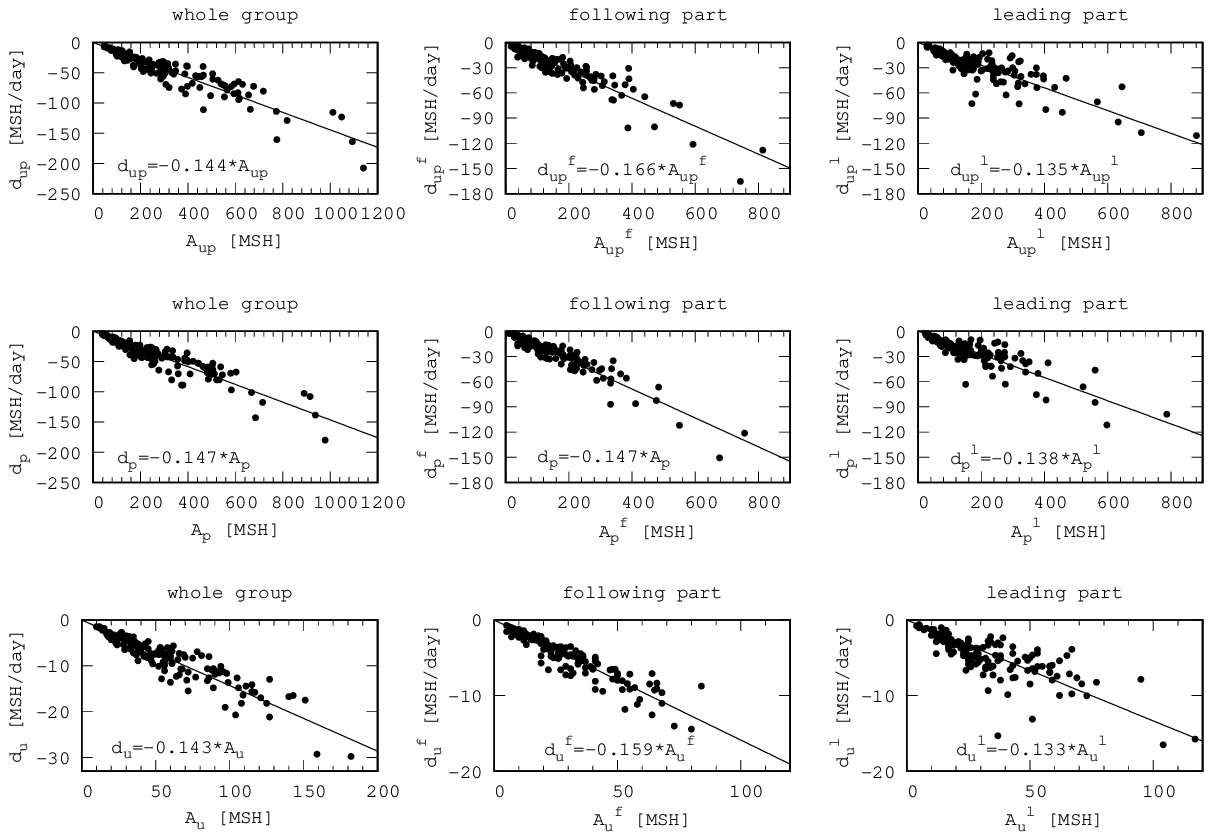}\\
\includegraphics[angle=0, width=0.9\linewidth]{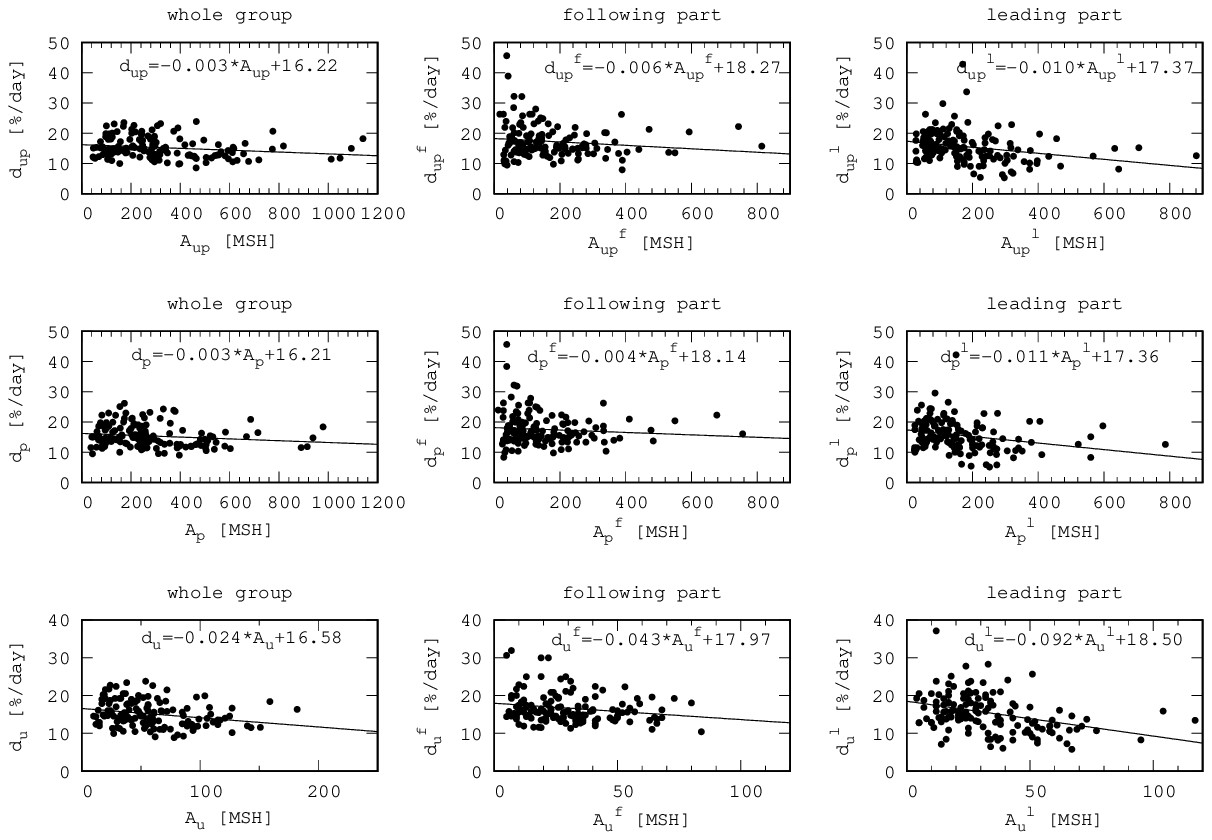}
\caption{The top three rows of figures show the daily decomposition of sunspot’s area depending on the proper areas measured in MSH, while the bottom three rows of figures depict the daily decomposition of sunspots measured in percentages. Decay speeds of the whole sunspot groups (top rows in the triplets), the penumbral (middle rows) and umbral (bottom rows) regression speed versus the appropriate maximum area. The fitted linear functions determine the decay rate of the sunspot groups and their parts. These figures are based on SDD data.}
\label{fig:decrates_sdd}
\end{figure}

\begin{figure}[h!]
\includegraphics[angle=0, width=0.9\linewidth]{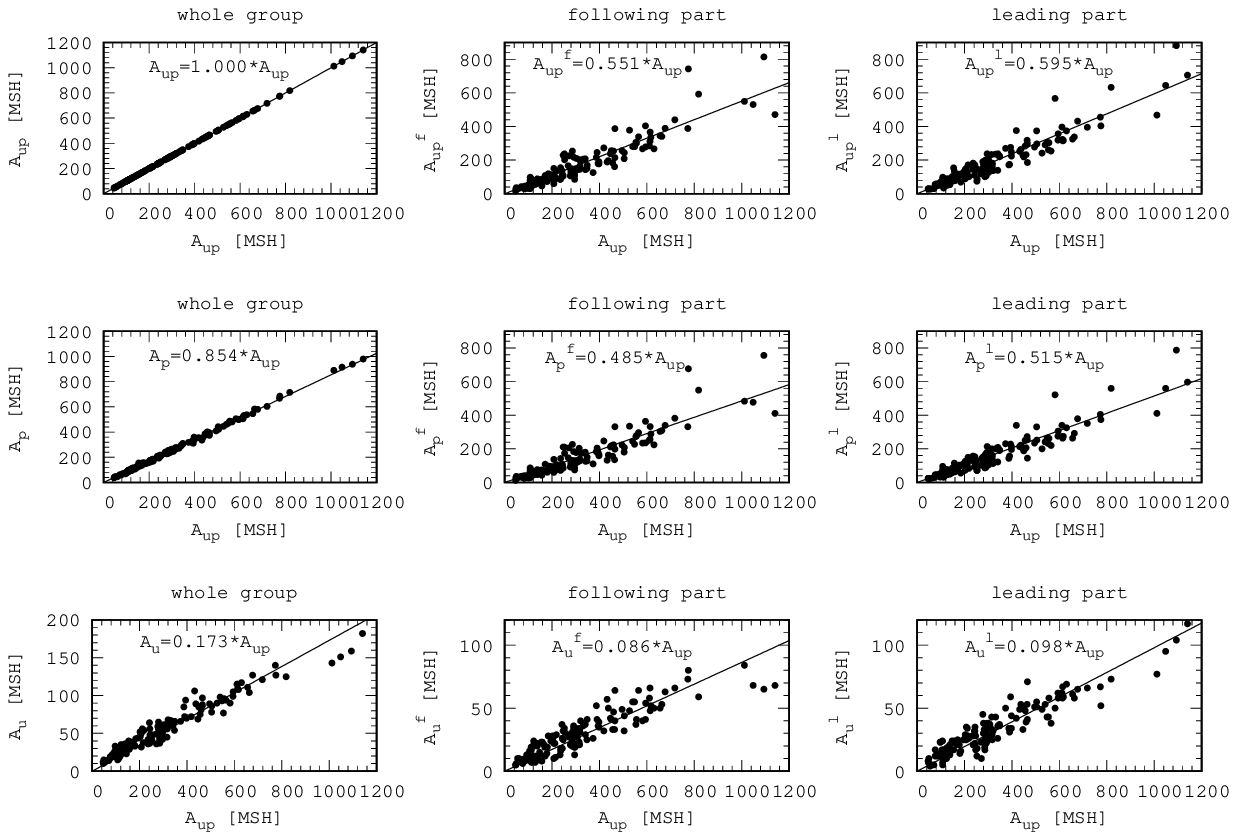}\\
\includegraphics[angle=0, width=0.9\linewidth]{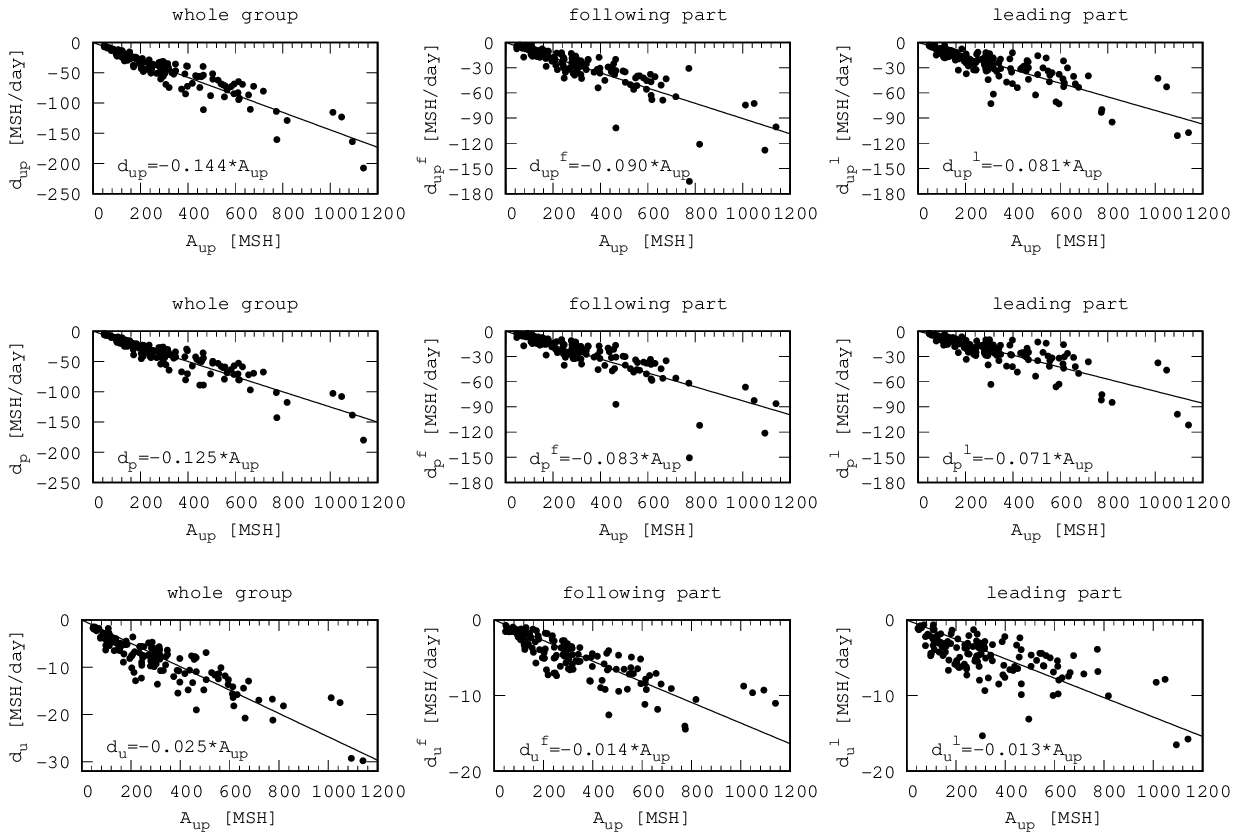}
\caption{Dynamics of the decay of sunspot groups and their subgroups of opposite magnetic polarities (leading and following umbrae and penumbrae). Panels of the upper three rows: areas of sunspot group’s parts depending on the total area of sunspot groups. Panels of the lower three rows: Decay speeds of the sunspot group’s parts depending on the total area of sunspot groups. 
Left column: whole group, middle column: following part, right column: leading part. Upper rows in the triplets: umbral and penumbral areas, middle rows: penumbral areas, bottom rows: umbral areas of sunspot groups respectively.}
\label{fig:cuprate}
\end{figure}

To study the details of the sunspot group evolution higher resolution is necessary. The SDD catalog makes this possible because it contains the magnetic data of the sunspots and the opposite polarity parts can be distinguished. Thus one sunspot group may have three maximum times (T) and three last areas ($a^w$), i.e., for the total sunspot group, its leading and following parts. This also means three different regression speeds (see Eq.~\ref{eq:decay}). Fig.~\ref{fig:decrates_sdd} illustrates the decay rates of the sunspot groups (left column) and well as their following (middle column) and leading parts (right column). The calculation of decay rates are made for the total area of sunspot groups (top panels), their penumbral area (middle panels) and umbral area (bottom panels). The percentage decay rates has also been calculated (bottom 9 panels of Fig.~\ref{fig:decrates_sdd}) and show the previously detailed property, i.e., the percentage decay rates depend also linearly on the proper maximum area. The higher the area of sunspot groups or their parts the lower the percentage decay rates. \\
During the decay the decrease of the area is not independent of the maximum area as one could expect from the linear trends of the upper three panels but it tends to be slower in large areas.

As Figs.~\ref{fig:decrates} and \ref{fig:decrates_sdd} illustrate, the magnitudes of the decay rates are independent of the applied data sets.

Fig.~\ref{fig:cuprate} compares the regression speeds of the parts of sunspot groups. These speeds depend both on the areas of the parts and the total area of sunspot groups. This means that the maximum total area of the sunspot group may determine the sizes of the leading and following parts (upper 9 panels) and their regression speeds (lower 9 panels). This means that knowing only the maximum total area of the sunspot group it can be determined how large can be its leading and following parts and what can be their regression speeds. The top 9 panels of Fig.~\ref{fig:cuprate} compare the areas of different parts of sunspot groups to the total equilibrium area of sunspot group (A$_{up}$). The maximum umbral area of sunspot group is about 17\%, while the maximum penumbral area is about 85\% of the total area. It can also be seen that the total area of the leading part is somewhat larger than that of the following part and the penumbral areas are about 5 times higher than the umbral areas.\\
The bottom nine panels of Fig.~\ref{fig:cuprate} illustrate the dependence of the different decay rates on the maximum area of total sunspot groups, i.e., the internal (leading and following umbral and penumbral) decay dynamics. The umbral decay rate is about 20\% of the penumbral decay rates. The slope of the decay rate of the following penumbra is the highest (-0.083 MSH/day /MSH) and that of the leading umbra (-0.013 MSH/day/MSH) is the lowest. This is statistical corroboration of the well known experience that the leading sunspots and their umbrae live longer.
Thus the usual process of the sunspot group's decay is the following. Firstly the following penumbrae vanish, then the leading ones. After that the following umbrae disappear earlier than the last leading umbrae.

\begin{figure}[h!]
\includegraphics[angle=0, width=\linewidth]{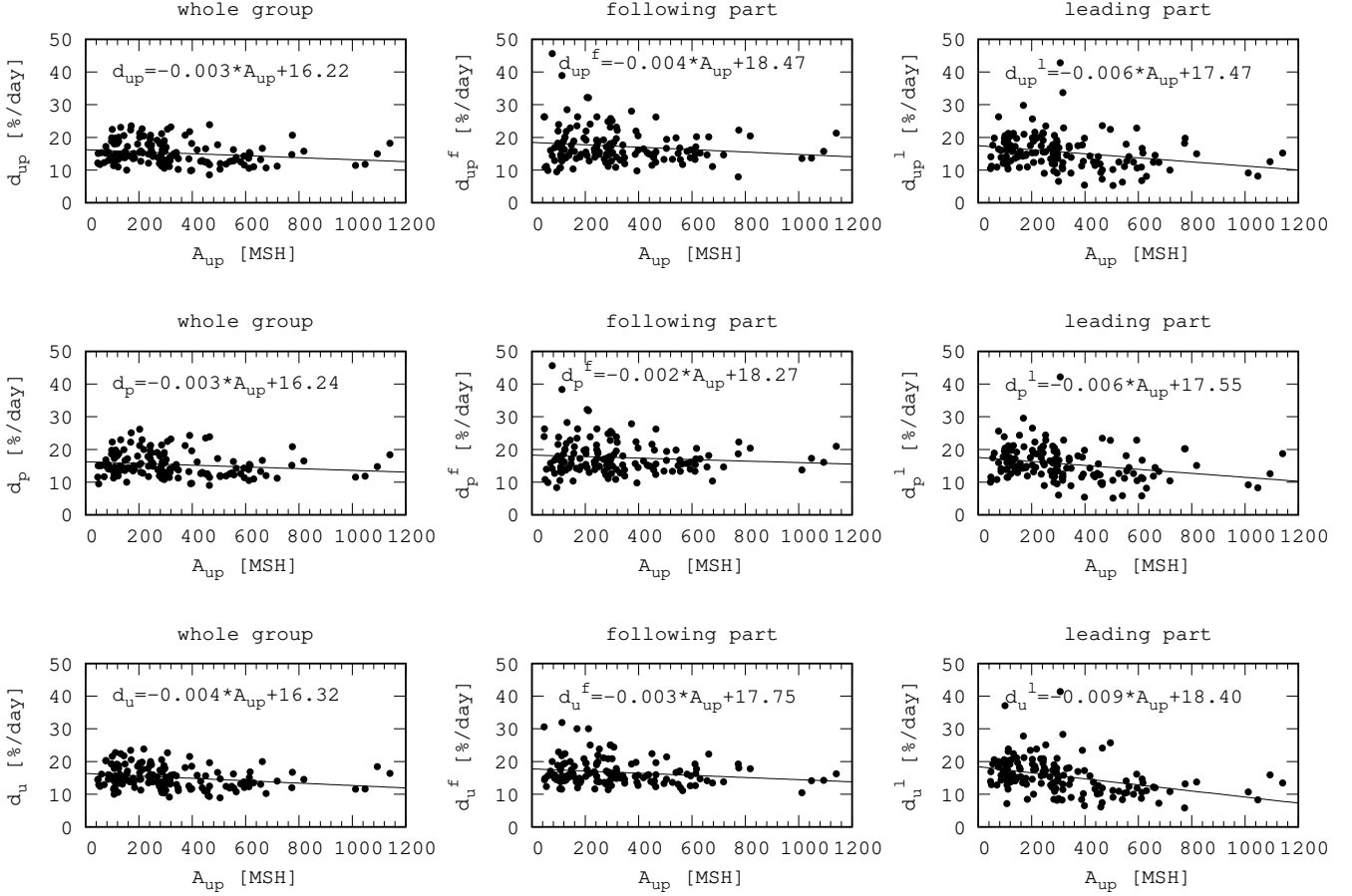}
\caption{The decay of the sunspot groups measured in percentage and its comparison to the total area of sunspot groups. Top panels: decay of the whole sunspot groups and their parts depending on the total area of whole sunspot groups. Middle panels: decay of the penumbrae and their parts depending on the total area of whole sunspot groups. Bottom panels: umbral decay depending on the total area of whole sunspot groups. Left column: whole group, middle column: following part, right column: leading part.}
\label{fig:cuprateszazcup}
\end{figure}

Without the dependence of the percentage decay rate on the area of sunspot group, the process of the decay should be like the earlier results (see bottom panels of Fig.~\ref{fig:cuprate}).
Fig.~\ref{fig:cuprateszazcup} illustrates how the daily percentage decay speed of sunspot groups depends on the total area of the whole sunspot group. This is an other view of the decay process which describe how the decay speed depends on the total area of sunspot group instead of its maximum area. This figure is also made by using SDD data and it seems to contradict to Fig.~\ref{fig:cuprate}. The lines mark the fitted linear functions which describe how the decay process of sunspot groups is if their internal parts (e.g. umbrae and penumbrae of their leading and following parts) are taken into account. Here the decay rate of the sunspot groups is not the weighted average of the leading and following parts because the decay rate is concerning to the proper maximum area, i.e., maximum area of the leading/following part.
However, it is difficult to survey the information content of Fig.~\ref{fig:cuprateszazcup} but from its data the schedule of disappearance of the sunspot groups can be illustrated by choosing a specific total area A$_{up}$ = 600 MSH. The slopes of the decay rates of the components are as follows:\\
following penumbra: 18.27 - 0.002*600  = 17.07 \%/day\\
following umbra: \hspace{0.40cm} 17.75 - 0.003*600  = 15.95 \%/day\\
leading penumbra:\hspace{0.28cm}  17.55 - 0.006*600 = 13.95 \%/day\\
leading umbra:   \hspace{0.70cm} 18.40 - 0.009*600  = 13.00 \%/day.\\
Although, the two results are not exactly the same but the main rhythm of the decay is the same. In both cases the leading umbrae have the longest lifetime because of the slowest decay, while the following penumbrae have the shortest life after their maximum area. As the percentage decay rate is more reliable because of the elimination of the maximum area the most probable process of the sunspot group's decay is when the size of the penumbrae decreases at first than that of the umbrae. The following part diminishes faster than the leading one.

\subsection{Cycle Phase Dependence}
The cycle phase dependency of the decay can be studied by using the modified GPR and DPD \citep{2016SoPh..291.3081B} covering 13 solar cycles. The study has been carried in a unified way, the lengths of all cycles have been divided to 10\% bins, the quantities have been averaged by each bin and these means have been averaged for the corresponding bins of all cycles.

\begin{figure}[h!]
\center
\includegraphics[angle=0, width=0.70\textwidth]{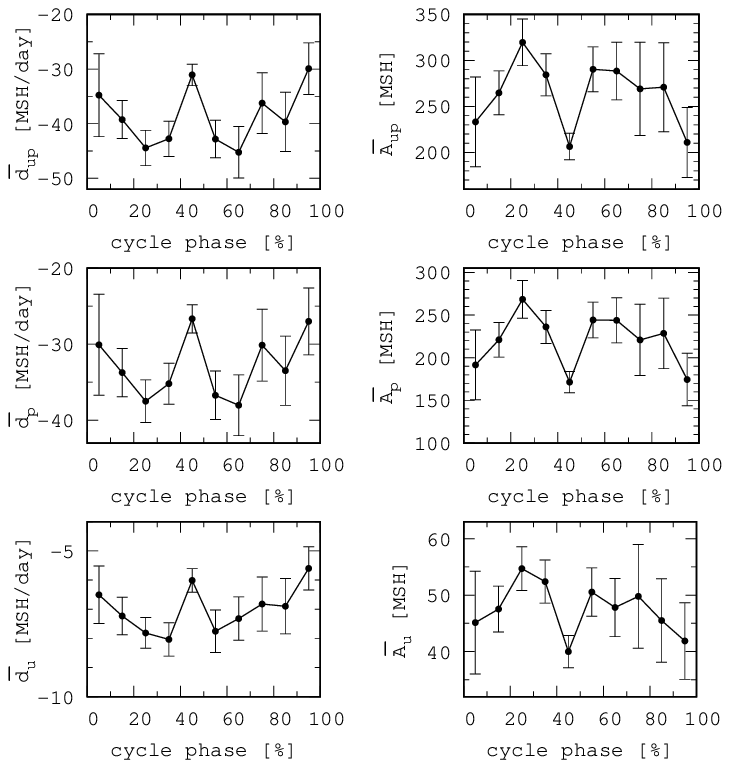}
\caption{Left panels depict the mean decay speeds of the whole sunspot groups (top), penumbrae (middle), and umbrae (bottom) averaged over 10 percentage bins of the 13 cycles. Right panels represent the averaged area of sunspot groups (top), penumbrae (middle), and umbrae (bottom) averaged over 10 percentage bins of the 13 cycles.}
\label{fig:cycleperc}
\end{figure}

The results are presented in Fig.~\ref{fig:cycleperc}, the left column shows the variation of the mean decay speeds in the course of the cycles, the right column depicts the variation of the areas, the rows contain the diagrams of total, penumbral and umbral areas of sunspots. It is conspicuous that the data of bin 40-50\% exhibit striking deviations from the trends of the diagrams, in these time intervals, during the maximum phase of the cycles, the mean decay speeds are definitely lower and the mean areas are smaller than in the 20-40\% and 50-70\% intervals, i.e., during the steepest development and decay phases of the cycles. The means of these quantities at cycle minimum (0-10\% and 90-100\%) are comparable to those at cycle maximum. The decreases of the mean decay speeds and the averaged areas of sunspot groups at the cycle maxima may be due presumably to the well-known Gnevyshev gap \citep{1967SoPh....1..107G}, the temporary depression of the solar cycle profile at its maximum. Their coincident decreases can follow from the strong linear relationship between the decay rates and the maximum areas of sunspot groups.

The cycle phase dependency of the decay has also been studied by \citet{2008SoPh..250..269H} but in lower resolution, they only compared the mean decay rates per sunspot in minimum and maximum. \citet{2011SoPh..270..463J} addressed higher resolution but on a yearly basis and his diagram does not show the salient data of Fig.~\ref{fig:cycleperc}.

\subsection{Cycle Dependencies and North--South Asymmetry}

\begin{figure}[h!]
\includegraphics[angle=0,  width=0.9\textwidth]{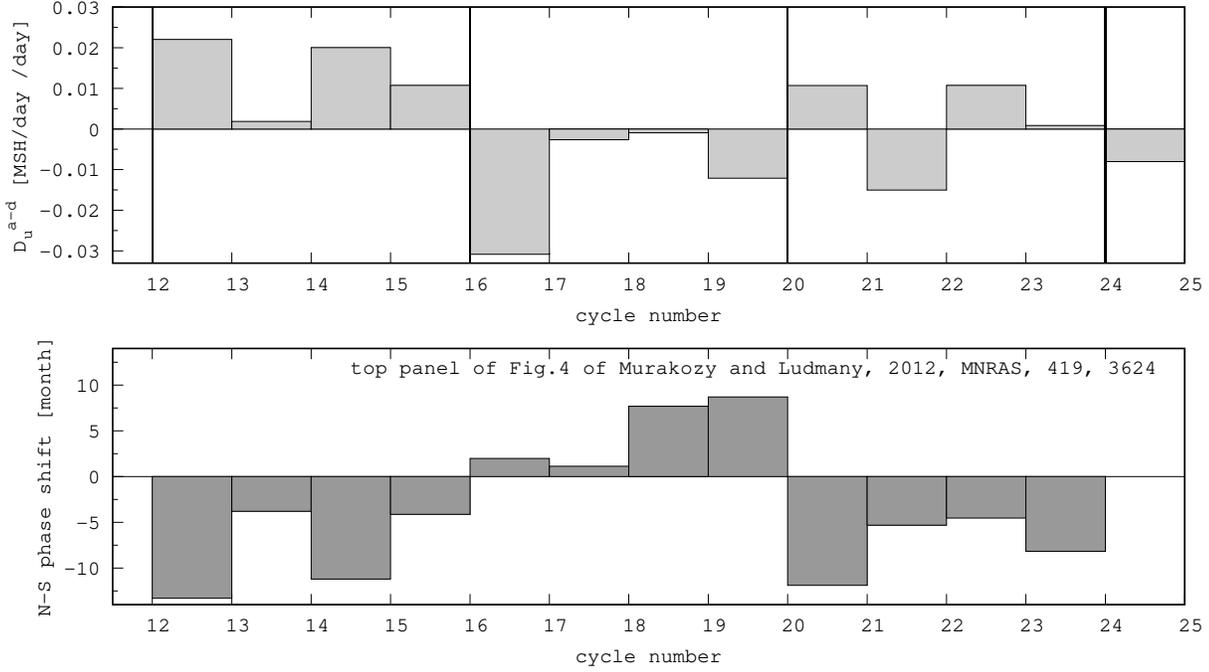}
\caption{Top panel: Umbral decay rates computed for the difference between the ascending and descending phases of cycles. The vertical black lines denote the boundaries of 4+4 cycles, i.e., the North-South asymmetry of solar cycles. The bottom panel shows solar N--S asymmetry detailed in \citet{2012MNRAS.419.3624M}.}
\label{fig:cycphwhole}
\end{figure}

As Fig.~\ref{fig:cycleperc} shows the decay rate is different before and after the maxima of cycles. Thus besides the cycle phase dependency of the sunspot decay a possible long-term cyclic dependency may also be addressed. Differences between the decay rates of umbrae in the ascending and descending phases of the cycles have been computed, they are plotted in the top panel of Fig.~\ref{fig:cycphwhole}.

These are calculated by using the umbral area of sunspot groups because the umbral area represents best the evolution of the emerged magnetic field.
As this figure illustrates the decay rates are higher in the ascending phases than in the descending phases during the first four cycles. This  difference changes its sign in the next four solar cycles. Between Cycles 20 and 23 the sign changed again and here the decay rates are higher in the ascending phases except Cycle 21. In the case of Cycle 24, the decay rates in the descending phase are higher, i.e., the difference changed its sign again.
The cyclic differences of the decay rates between the ascending and descending phases are surprisingly organized into 4+4 stripes, the only one exception is Cycle 21.

The variation exhibits an unexpected similarity to an earlier found pattern of 4+4 cycles in the North-South asymmetry \citep{2012MNRAS.419.3624M} in which the southern hemispheric cycle leads in four Schwabe cycles (12-15) and the northern hemispheric cycle takes over the leading role in the next four cycles (16-19). This regularity is cited from the paper of \citet{2012MNRAS.419.3624M} in the lower panel of Fig.~\ref{fig:cycphwhole}. 

\begin{figure}[h!]
\includegraphics[angle=0, width=0.9\textwidth]{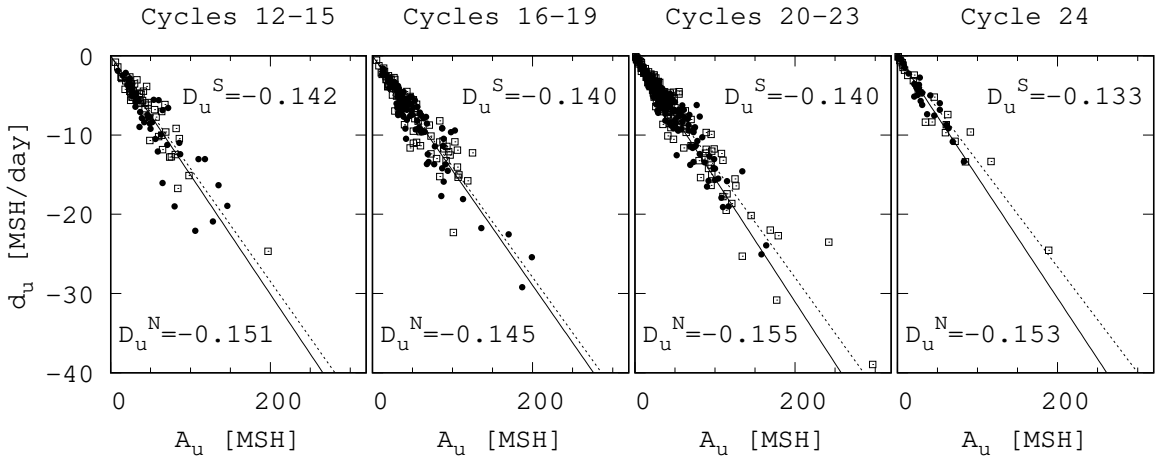}
\caption{The North-South umbral decay speeds depend on their maximum area. Their linear fitted function plotted with a solid line for the northern hemisphere, and with a dashed line for the southern hemisphere. D$_u$ values mark the steepnesses of the linear fitted functions, i.e., the umbral decay rates.}
\label{fig:NS}
\end{figure}

In contrast to the decay rates in the ascending-descending phases the North-South asymmetry of the decay rates does not exhibit the pattern of 4+4 Schwabe cycles as shown in Fig.~\ref{fig:NS} for all groups of 4 cycles. This means that a permanent hemispheric asymmetry exists in the decay process of sunspot groups. The dependencies of these hemispheric umbral decay speeds on the maximum umbral area during four Schwabe cycles are shown in the panels of Fig.~\ref{fig:NS}. It can be seen that the northern hemispheric umbral decay rate is always higher. Although Cycle 21 is the only exception, its hemispheric decay rates are as follows;  D$_u^N$=0.166 MSH/day /MSH and D$_u^S$=0.135 MSH/day /MSH. This is an unexpected result but it seems that there is an asymmetry between the solar northern and southern hemispheres on longer time scale as well.
\citet{2013SoPh..287..215M} obtained that the tilt angles of sunspot groups normalized to the latitudes are anti-correlated to the cycle strenght in the southern hemisphere but in the northern hemisphere this kind of relationship can not be pointed out.

The hemispheric average tilt angle is lower in the southern hemisphere based on study of data set 1917-1985.
\citet{2012ApJ...758..115L} pointed out that that the tilt angles in the southern hemisphere are higher with about 6\% than in the northern hemisphere and the difference between them increases toward higher latitudes.
\citet{2011SoPh..270..463J} investigated the growth and decay of sunspot groups on the period of 1874-2009 and found a 60-80 year cycle in the percentage decay rates. A North-South asymmetry was also pointed out in the percentage growth rate of groups and noted that this may be related to the 33-44 year cycle in the solar activity.

\subsection{Latitudinal Dependency}

\begin{figure}[h!]
\center
\includegraphics[angle=0, width=0.75\linewidth]{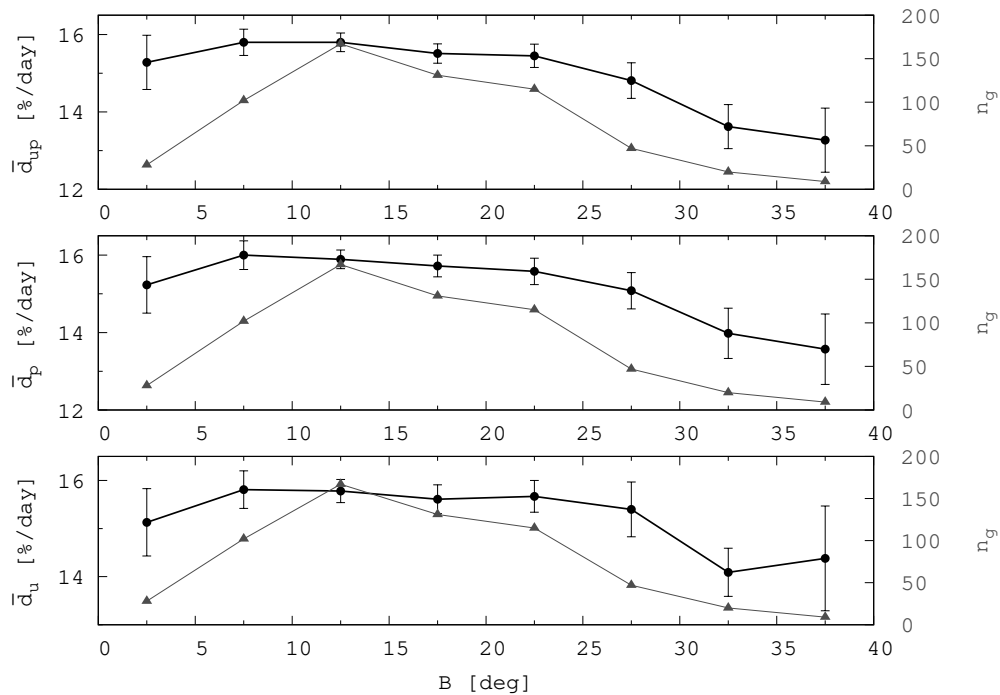}
\caption{Latitudinal distribution of the mean percentage decay rates of total sunspot groups (top panel), penumbrae (middle panel), and umbrae (bottom panel) plotted with black dots and thick black lines. The average values with their standard errors are calculated in latitudinal belts of 5 degrees, their values are marked on the left axis. The sunspot group number counted in each belt denoted with grey triangles and their values are scaled on the right axis.}
\label{fig:bdep}
\end{figure}

The latitudinal distribution of the decay rates may also yield information on the conditions affecting the sunspot decay. The average of the percentage decay speeds and their standard errors are calculated in latitudinal belts of 5 degrees without distinction of the hemispheres, the results are plotted in Fig.~\ref{fig:bdep} along with the numbers of sunspot groups in each belt.

The trends are similar for the three types of area data, the decay is more or less uniform between the latitudinal belts of 7 and 27 degrees and it is lower close to the equator and at highest latitudes, i.e., around cycle minima. Latitudinal dependency of the sunspot decay speed was noted earlier by \citet{1995SoPh..157..389L} and \citet{2008SoPh..250..269H} but they found that the higher latitude spots decay more rapidly than the lower latitude spots.
Fig.~\ref{fig:bdep} also shows that the decay rates are substantially lower toward higher latitudes than in the middle and low latitude bins. This is similar to the findings of \citet{2012ApJ...758..115L} who investigated the tilt angles of more than 36,000 sunspot groups and found a sharp deviation of Joy's law at higher latitude bins.

\section{Summary and Conclusion}
The aim of the present study is to reveal those conditions that may modify or contribute to the sunspot decay process. The results and the conjectured implications can be summarized as follows.

(i) The decay of sunspot groups is linear as a function of the maximum total area calculated in total, umbral, and penumbral areas alike. The percental decay rates of larger groups are smaller this may explain their longer lifetime, see Figs.~\ref{fig:decrates} and \ref{fig:decrates_sdd}.

(ii) The results are similar for both the leading and following parts of the groups by using the SDD sunspot database containing magnetic data as well. The decay rates of areas are in descending order: following penumbrae, following umbrae, leading penumbrae, leading umbrae (see Fig.~\ref{fig:cuprateszazcup}). The differences between umbral an penumbral decays may imply that their mechanisms extend to different depths, the penumbral decay is controlled by processes at the surface while the umbral decay progresses under the impact of the velocity fields of the deeper layers. The leading-following difference may be the consequence of the greater compactness of the leading subgroup \citep{2014SoPh..289..563M}. About the role of the penumbra during the decay \citet{2014ApJ...785...90R} argue that the penumbra could stabilize the near surface layers against decay and delays the fragmentation in the subsurface layers. This difference could also be caused by the convective intrusion into the penumbra at the same depths (this theory, i.e., role of the U/P rate in the decay will be checked in a separate paper).

(iii) The area at the maximum state of the sunspot group determines its subsequent life. By knowing the highest area of the sunspot group, the highest leading and following areas of the umbrae and penumbrae as well as their decay rates can be estimated (see Fig.~\ref{fig:cuprate}).

(iv) A strange cycle phase dependency of the decay can be pointed out. In the ascending phase of the cycle the mean sunspot group area increases and the mean decay rate also increases and these trends reverse in the descending phase. However, around the maximum between about 40-50\% of the cycle length, an abrupt deviation can be detected from this trend (see Fig.\ref{fig:cycleperc}). 
This result seems to be related to the Gnevyshev gap \citep{1967SoPh....1..107G}, the transitory depression of the solar cycle profile at its maximum where the mean sunspot group area also decreases. Here the decrease of the decay rate may be related to its connection to the sunspot group area.

(v) The differences between the decay rates of ascending and descending phases of the solar cycles exhibit a north-south asymmetry that changes sign by four Schwabe cycles. This variation is similar to the earlier found pattern of 4+4 cycles in the phase lags of hemispheric cycles \citep{2012MNRAS.419.3624M}, see Fig.~\ref{fig:cycphwhole}. Both regularities are fairly enigmatic but their similarity raises the question whether they are caused by some still unidentified variation.

(vi) In the latitudes of the activity belt, there is a slight latitude dependence in the values of percental decay rates (see Fig.~\ref{fig:bdep}). The higher the latitude the lower the decay rates. The conjectured background may be similar to that mentioned in point (ii), the magnetic flux ropes may extend to different depths at high and low latitudes therefore they may be exposed to different degrees to the turbulent erosion exerted on the magnetic field by the surrounding velocity field. Another finding is that the higher latitude decay rates are substantially lower than those at lower latitudes. This can hint at some differences between the circumstances at higher and lower latitudes similarly to the result of \citet{2012ApJ...758..115L} on the latitudinal dependence of tilt angles.

The raised theoretical suggestions will be studied in a separate paper.

\section*{Acknowledgements}
The author is indebted for the received funding from National Research, Development and Innovation Office -- NKFIH, 129137. Thanks are due to Dr. Andr\'as Ludm\'any for reading and discussing the manuscript and the unknown referee for her/his comments and suggestions make the article easier to understand.





\bibliography{Murakozy_biblio}{}
\bibliographystyle{aasjournal}

\end{document}